\documentclass[11pt,a4paper,twoside]{article}

\usepackage[utf8]{inputenc}
\usepackage[margin=1in,footskip=0.35in]{geometry}
\usepackage{amsmath,amssymb}

\usepackage[super]{cite}
\usepackage{graphicx,caption}
\usepackage{xcolor}
\usepackage[hidelinks]{hyperref}
\usepackage{fancyhdr}

\begin{document}
\noindent
Modern Physics Letters A, Vol.\ \textbf{39}, No.\ 12, 2450053 (2024)\footnote{This manuscript is also found with comments on  \href{http://micro.cozmo.dk}{www.micro.cozmo.dk}.}\\
\href{https://doi.org/10.1142/S0217732324500536}{DOI: 10.1142/S0217732324500536}

\vspace{2cm}

\begin{center}
\textbf{\Large Micro-cosmos model of a nucleon}\\ \

\vspace{1cm}
Michael Cramer Andersen\\ \

\emph{Christianshavns Gymnasium, Prinsessegade 35, DK-1422 Copenhagen, Denmark}\\
\emph{micran@gmail.com}\\ \

Received (27 October 2023)\\
Revised (24 March 2024)\\
Accepted (2 April 2024)\\
Published (4 May 2024)
\end{center}



\vspace{1cm}
\begin{abstract}
This study explores the age-old quest to construct a geometric model of a quantum particle. While static classical particle models have largely been dismissed, the focus has now shifted to intricate dynamic models that hold the promise of reconciling general relativity with quantum mechanics.

We propose that matter particles can be described as radiation confined within dynamically curved spacetime regions, without the need for quantization of space and time, and using standard field equations and natural Planck units. Specifically, we investigate a cyclic or oscillating radiation-dominated micro cosmos undergoing repeated bouncing.

Our methodology employs integration, with carefully defined initial conditions. The results include several observable properties characteristic of quantum particles. We calculate the total mass, revealing a compelling inverse proportionality between mass and radius identical with the de Broglie relationship. Applying this model to protons, we discover a profound and surprisingly simple relationship between the proton's radius and mass expressed in Planck units. This enables a definition of the proton radius that aligns remarkably well with the 2018 CODATA value.
Furthermore, our analysis demonstrates that the radial density profile of the proton (or nucleon), averaged over a cycle time, increases toward the center. 
The problem of embedding the micro cosmos within a background spacetime is also described.

These results underscore the relevance of general relativity in the domain of nuclear physics. Moreover, the model offers a fresh perspective that can stimulate new ideas in the ongoing quest to unify general relativity with quantum physics.\\


\end{abstract}

\newpage
\pagestyle{fancy}
\fancyhead{} 
\renewcommand{\headrulewidth}{0pt}
\fancyhead[RE]{\emph{Micro-cosmos model of a nucleon}}
\fancyhead[LO]{\emph{M.C.\ Andersen}}

\section{Introduction: The particle problem in general relativity}
This study follows the tradition of a geometrical description of quantum particles.
Einstein had the dream of constructing a model of particles as static point-like concentrations of electromagnetic and/or gravitational fields, and that singularities were avoided.
Early attempts included the Einstein-Rosen bridge/wormhole (1935) and Wheeler's Geons (1955), both of which were far from describing known particles.

The Kerr-Newman metric (1965) describing a rotating black hole with electrical charge was mentioned by Carter\cite{Carter1968} to have some similarities with a Dirac-electron.

The \emph{strong gravity} (or bi-scale) approach in the 1970s introduced models of hadrons constructed by solutions to Einstein's field equations, used with success in astrophysics and cosmology, just on a much smaller scale: micro black holes \cite{Sivaram-Sinha1977,Sivaram-Sinha1979} and cyclic micro universes \cite{CaldirolaPavsicRecami1978,CaldirolaRecami1979,ItalianoRecami1984}.
There is a striking similarity between a quantum particle and a mini-black hole. Both objects are described by the same three physical quantities (owing to the fundamental space-time symmetries): mass, electrical charge, and angular momentum (spin). The fact that both objects are assumed to be (nearly) point-like makes the rotating charged black hole solution an interesting candidate for a space-time model of quantum particles.
 
A serious problem with modeling a quantum particle using a black hole is, that the size of the horizon of a typical hadron is significantly smaller than the actual size of the particle. The Schwarzschild radius of a proton, $r_s(m_\mathrm{p})=2.5\cdot 10^{-54}$ m, is a factor $\sim 10^{38}$ too small and smaller than the Planck length.
A method to work around this problem was investigated in strong gravity with Salam as one of the originators\cite{Salam1972,SalamSivaram1993}.
The strong gravity field was described by Einstein's field equations for gravity. However, $\kappa_g=8\pi G/c^4$ was replaced with a strong coupling constant $\kappa_f=10^{38}\cdot \kappa_g$, thus moving the energy scale of the theory to the hadron scale. The horizon of the black hole with the mass of the hadron then fitted the experimental value (within an order of magnitude). The hope was that problems in hadron physics could be translated into general relativity, and that exact solutions could be found that could be translated back into hadron physics.
An entire phenomenology was developed, including a de Broglie-like relation, a strong nuclear force similar to the Yukawa potential, point-like particles inside the hadron (quarks), confinement, asymptotic freedom, and a particle spectrum of hadrons. In one study\cite{RecamiZanchin1994} it was shown that stable non-evaporating Kerr-Newman black holes in a de Sitter universe was exhibiting Regge-like relations between $m$, $a$ and $Q$, similar to the relations found for quantum particles. However, the models had little impact on particle physics.

The problem of understanding the stability of elementary particles was studied further with the development of a Kerr-Newman model of the Dirac-electron. Burinskii\cite{Burinskii2008,Burinskii2015} found that the Dirac equation was ``a master equation controlling the motion and orientation of the Kerr-Newman twistorial space-time structure''.

In the \emph{graviweak unification} approach\cite{Onofrio2013,Onofrio2014} it is conjectured that the \emph{weak interaction}, expressed by the Fermi constant $G_F$, is related to Newton's constant $G_N$: $G_F=\xi\cdot(\hbar/c)^2\cdot G_N$, where $\xi= 1.74\cdot 10^{33}$ is a dimensionless constant which is absorbed if $G_N$ is renormalized at the attometer scale which is possible in $f(R)$ gravity theories. In this framework weak interactions are ``short-distance gravitational phenomena'', $G_N$ is running and at the attometer scale, $G_F$ is an effective $G_N$ correcting for quantum fluctuations of the spacetime foam (such as wormholes).
Renormalizing $G_N$ by a factor of $10^{33}$ increases the Schwarzschild horizon of the fundamental particles\cite{Onofrio2014}, which is analogous to strong gravity.

The main reason why strong gravity has been largely disregarded is the assumption of a special coupling constant for hadrons, which appears ad hoc. 
This criticism must also apply to graviweak unification. The mix of concepts from general relativity and particle physics bridging the two separate disciplines was challenging in the 1970s. Today, when the AdS/CFT correspondence is used to translate problems with strong coupling in nuclear physics into problems with weak gravitational coupling in string theory, it is essentially a continuation of the technique used in strong gravity. The obvious similarity in the phenomenology of fundamental particles and astrophysical objects is an indication that they can be described by the same underlying physics.

Even if a (static) black hole alone cannot describe a quantum particle, dynamic models including a micro cosmos should be investigated further.

\section{Method of the study}
The overall research question is as follows: Is it possible to construct a realistic dynamic geometric model of a quantum particle? Our model is inspired by the strong gravity approach but instead of modifying the coupling constant in Einstein's field equations, the model is based on the standard field equations and natural Planck units, which are appropriate for describing fundamental phenomena, such as quantum particles. General relativity is assumed to be valid for distances $l\gtrsim l_\mathrm{Pl}$ and densities $\rho\lesssim \rho_\mathrm{Pl}$. The hadrons and leptons lie within this regime. The quantization of spacetime is not relevant other than at the Planck scale, and possibly during the bounces. Planck units with $c$, $G$, and $\hbar$ were used.

The fact that quantum particles are connected to electromagnetic radiation, as observed in the pair creation and annihilation processes, motivates the hypothesis that matter particles consist of radiation trapped in curved dynamic spacetime regions, such as a cyclic micro cosmos. The mass of nucleons is dominated by the gluon field which is closely related to electromagnetic radiation.

The remainder of the paper is structured as follows.
First, the Friedmann equation for a closed radiation-dominated model universe is defined, and the solution for the scale factor is rewritten in geometric units (Section \ref{SectionCyclicMicroCosmos}).
The model was applied to calculate the time-varying mass and total mass integrated over one cycle, and the mass and radius were found to obey an inverse proportionality identical to the de Broglie relation describing the wave-particle duality (Section \ref{SectionMassMicroCosmos}).
Next, the model was applied to the proton, and the observable quantities of mass and radius were compared with experimental values (Section \ref{SectionModelProton}). A hitherto unknown factor of order unity was discovered to be $2/\pi$ enabling an exact definition of the proton radius.
The average density and radial density profile were calculated, and applications in nuclear physics were discussed (Sections \ref{SectionNucleonDensity}-\ref{SectionNuclearPhysics}).
The electron is discussed (Section 9).
Finally, the problem of a point-like phase during the bounces and the problem of embedding are described (Sections \ref{SectionSingularity} and \ref{SectionEmbedding}). Future generalizations of the model are proposed (Section \ref{SectionGeneralization}).
Some calculations are provided in the appendices.

\section{Theoretical framework: The cyclic micro cosmos}\label{SectionCyclicMicroCosmos}
Assume that a particle can be described by a spherical bag of radiation with average density $\rho_\mathrm{particle}$. This entity sits in a background metric that is locally approximately flat with an average density $\rho_\mathrm{background}\ll \rho_\mathrm{particle}$. 
The average density of our Universe went below the nuclear density of $\sim 10^{18}$ kg/m${}^3$, around the era of nucleosynthesis beginning a few seconds after Big Bang. Before that time the boundary between particle and host universe was not well defined due to density fluctuations and possibly metric fluctuations of comparable magnitude as $\rho_\mathrm{particle}$.\\
\indent The change in $\rho_\mathrm{background}$ is assumed to be slow so that the background metric may be regarded as stationary and without impact on the particle within the relevant dynamical time scale of the particle ($\Delta t\lesssim 10^{-23}$ s for nucleons).
The question of how the particle fits with the background metric is described in Section \ref{SectionEmbedding}.   

Consider a radiation-dominated closed homogeneous and isotropic model universe
with scale factor $a(t)$, global curvature $k=1$ comoving radial distance $r$ and
radiation (mass) density $\rho_\mathrm{r}=\rho_\mathrm{r,0}(a_0/a)^{4}$, where $a_0=l_\mathrm{Pl}$ is the minimum length scale and $\rho_\mathrm{r,0}$ is the density at a given time $t_0$.
The total density parameter is $\Omega_0\equiv\Omega_{r}=\rho_\mathrm{r}/\rho_\mathrm{crit}$, with the critical density $\rho_\mathrm{crit}=3H^2/8\pi G$. The spatial curvature density is written: $\Omega_{k}\equiv 1-\Omega_0$, the Hubble parameter $H\equiv\frac{1}{a}\frac{da}{dt}$ and the Hubble constant is $H_0$ at $t_0$.

The dynamics is described by the Friedmann equation (in cosmological units):
\begin{equation}\label{Friedmann-Radiation-Cosmological}
\frac{H^2}{H_0^2}=\Omega_{0} \left(\frac{a_0}{a}\right)^4 +(1-\Omega_0)\cdot\left(\frac{a_0}{a}\right)^2
\end{equation}
We want to eliminate $\Omega_0$ and introduce the \emph{maximal size}, $a_\mathrm{max}=\sqrt{\Omega_0/(\Omega_0-1)}\cdot a_0$, which is reached when $H=0$ and it is used to eliminate $\Omega_0=a_\mathrm{max}^2/(a_\mathrm{max}^2-a_0^2)$.

The resulting Friedmann equation is:
\begin{equation}\label{Friedmann-Radiation-Geometrical}
\left(\frac{da}{dt}\right)^2=H_0^2\cdot\frac{a_\mathrm{max}^2\cdot a_0^4}{a_\mathrm{max}^2-a_0^2}\cdot\left(\frac{1}{a^2} -\frac{1}{a_\mathrm{max}^2}\right)
\end{equation}
It is convenient to introduce the \emph{cycle time}, defined as the duration between two bounces, which is determined by the condition $a(t_\mathrm{crunch})=a_0$:
\begin{equation}\label{Tcrunch}
t_\mathrm{crunch}
=\frac{2}{H_0}\cdot\frac{a_\mathrm{max}^2-a_0^2}{a_0^2}
\end{equation}
Equation (\ref{Friedmann-Radiation-Geometrical}) is solved by separation of the variables $t$ and $a$ and integration (with $t_\mathrm{0}=0$), 
resulting in an expression for $t(a)$, where $H_0$ is also eliminated and for $a(t)$:
\begin{equation}\label{Time-scalefactor}
t(a)=t_\mathrm{crunch}\cdot
\left(\frac{1}{2}\pm\frac{1}{2}\cdot\sqrt{\frac{a_\mathrm{max}^2-a^2}{a_\mathrm{max}^2-a_0^2}}\,\right)
\end{equation}
\begin{eqnarray}
a(t)&=&
\sqrt{a_0^2+4\cdot(a_\mathrm{max}^2-a_\mathrm{0}^2)\cdot
\left(\frac{t}{t_\mathrm{crunch}}\right)
-4\cdot(a_\mathrm{max}^2-a_\mathrm{0}^2)\cdot
\left(\frac{t}{t_\mathrm{crunch}}\right)^2
}\label{ScalefactorGeometricExact}
\end{eqnarray}
The scale factor oscillates between $a(0)=a(t_\mathrm{crunch})=a_0$ and $a(\frac{1}{2}\!\cdot\! t_\mathrm{crunch})=a_\mathrm{max}$ making the curve a curtate trochoid with no singularities during the bounces (if $a_0=0$ it is a common cycloid).

When the micro cosmos is embedded in a larger nearly flat spacetime and integrated over one cycle time, it will be indistinguishable from a particle. The embedding procedure is described in Section \ref{SectionEmbedding}. The spacetime model can now be applied to particle physics. The observable quantities of the mass, radius, and density are calculated in the following sections.

\section{Mass of the micro cosmos and quantum physics}\label{SectionMassMicroCosmos}
The \emph{mass} of the radiation filled \emph{micro cosmos}, $m_\mathrm{mc}$, is found, relying on the equivalence between energy and mass, by multiplying density, $\rho_r$\,=\,$\rho_{r,0}(a_0/a(t))^{4}$,
and volume, $V$\,=\,$\frac{4\pi}{3}a(t)^3$ inserting $a_0=l_\mathrm{Pl}$ and $\rho_{r,0}=m_\mathrm{Pl}/(\frac{4\pi}{3}l_\mathrm{Pl}^{3})$:
\begin{eqnarray}\label{MicroMassTimevarying}
m_\mathrm{mc}(t)
=\rho_r\cdot V=
\frac{l_\mathrm{Pl}\cdot m_\mathrm{Pl}}{a(t)}
\end{eqnarray}
The mass is time-varying and inversely proportional to the length scale $m\propto 1/a$, similar to the de Broglie relation, which sets the scale where the wave properties of any quantum particle emerge, or the related (and smaller) \emph{Compton wavelength}, which sets the scale for describing a quantum particle as a single point-like entity:\footnote{If a particle is probed at distances smaller than $\lambda_\mathrm{C}$ (corresponding to a higher momentum of the probe), QED predicts that pairs of virtual particles and antiparticles emerge, contributing to the mass of the particle and the screening of the charge.}
\begin{equation}\label{ComptonPlanck}
\lambda_\mathrm{C}=\frac{\hbar}{m\cdot c}
=\frac{l_\mathrm{Pl}\cdot m_\mathrm{Pl}}{m}
\end{equation}
Equation (\ref{MicroMassTimevarying}) is identical to the Compton relation if $\lambda_\mathrm{C}$ and $a(t)$ are comparable in size. In the middle of a cycle, $a(\frac{1}{2}t_\mathrm{crunch})=a_\mathrm{max}$ indicating that $a_\mathrm{max}=\lambda_C$ during maximum (observed from inside the micro cosmos).

When the radiation-dominated micro cosmos, respectively, expands or contracts by a factor $N=a_\mathrm{max}/a_0$, the radiation experiences a cosmological \emph{redshift} or \emph{blueshift} and the mass decreases or increases by a factor $N$.
The scale factor (and thereby the mass) changes rapidly between $a_0$ and $a_\mathrm{max}$ and the properties of the micro cosmos changes rapidly between the two states, making it a good candidate for a classical object with built-in wave-particle duality:
\begin{itemize}
\item \emph{Extended phase (wave):} When $a(t)\lesssim a_\mathrm{max}$ the object has wave properties -- it can interfere with other objects and has an insignificant mass.
\item \emph{Point-like phase (particle):} When $a(t)\gtrsim a_0$ the object has particle properties -- it is localized and has a large mass.
\end{itemize}
When the micro cosmos particle is probed at smaller length scales, the mass increases.
Although the mass changes rapidly within each cycle, the particle still exists continuously throughout the cycles.
Averaged over time scales $t\gg t_\mathrm{crunch}$ the micro cosmos will have a constant concentration of energy within a finite volume.

The \emph{total mass}, still seen from the inside, is found by integration of $m_\mathrm{mc}(t)$ from 0 to $t_\mathrm{crunch}$ normalizing over one cycle (see appendix \ref{AppendixIntegrationMass}):
\begin{equation}\label{MicroMassTotalInner}
m_\mathrm{mc}
=\frac{1}{t_\mathrm{crunch}}\int^{t_\mathrm{crunch}}_0 \hspace{-3mm}m_\mathrm{mc}(t)\,dt
=\frac{\pi}{2}\cdot\frac{a_0}{a_\mathrm{max}}\cdot m_\mathrm{Pl}
\end{equation}
From outside the micro cosmos, a particle with a certain mass and radius is observed.
The maximal radius, $a_\mathrm{max}$, is an intrinsic parameter in the curved (compactified) micro cosmos.

During a full oscillation of the micro cosmos ($\Delta t=2\cdot t_\mathrm{crunch}$), a test particle moves a distance of $4\cdot a_\mathrm{max}$ (from $+a_\mathrm{max}$ through the diameter twice and back to the starting point at $+a_\mathrm{max}$). Observed from the outside, this distance corresponds to a rotation around the particle, $l_\mathrm{particle}$, but it is not exactly the circumference because the spacetime moves in and out in a wavy curve. To shift from the inner radius to an outer measure of the size ('wavelength') of the particle, we shift $a_\mathrm{max}\to\frac{1}{4}\cdot l_\mathrm{particle}$ in equation (\ref{MicroMassTotalInner}) and find
the outer \emph{total mass}:
\begin{equation}\label{MicroMassTotalOuter}
m_\mathrm{mc}
=2\pi\cdot\frac{\hbar}{l_\mathrm{particle}\cdot c}
=\frac{h}{l_\mathrm{particle}\cdot c}
\end{equation}
The reduced Planck constant, $\hbar=h/2\pi$, is used to simplify expressions involving angular momentum, which is associated with intrinsic rotation or oscillation. The oscillation of the micro cosmos is just such an intrinsic oscillation.

The size is equal to the de Broglie wavelength with $m_\mathrm{particle}=m_\mathrm{mc}$ and $v=c$:
\begin{equation}\label{SizeParticle}
l_\mathrm{particle}
=\frac{h}{m_\mathrm{particle}\cdot c}
=\lambda_\mathrm{dB}
\end{equation}
We have not restricted us to a specific particle.
In principle, any quantum particle should be considered but the geometrical nature of the model demands a precise knowledge of the radius.

An electron would be a natural first choice to apply the model on. But its radius is only determined experimentally by a limit: $r_e\lesssim 10^{-18}$ m. Different characteristic radii, Compton length, and classical electron radius are derived from the mass and related by the fine structure constant $\alpha$ which requires a geometrical interpretation. The electron is discussed further in Section 9.

Instead, we will apply the model to a proton, which is the most stable and best-studied hadron with both experimentally well-defined mass and radius. In QCD the proton or nucleon is modeled by three valence quarks, a gluon field, and virtual sea quarks. More than 99 \% of the energy is associated with the gluon field which is a generalized short-range electromagnetic field possessing color charge. The total net color charge of the gluon field is white (as the total color charge of the constituting quarks), making the total gluon field very similar to electromagnetic radiation.

\section{A model of the proton: Mass and radius}\label{SectionModelProton}
The mass and radius of a proton expressed in Planck units, $r_\mathrm{p}/l_\mathrm{Pl}\cong m_\mathrm{Pl}/m_\mathrm{p}\cong 10^{19}$, satisfy
an inverse proportionality, except for a numerical factor $n_\mathrm{c}$ of order unity:
\begin{equation}\label{RadiusMass1019}
m_\mathrm{p}\cdot r_\mathrm{p}
=n_\mathrm{c}\cdot m_\mathrm{Pl}\cdot l_\mathrm{Pl}
=n_\mathrm{c}\cdot \frac{\hbar}{c}
\end{equation}
The reason why the two ratios are very close, but not exactly equal, is a mystery known as a large number coincidence. Carr and Rees\cite{CarrRees1979} expressed it through $\sqrt{\alpha_\mathrm{G}}$ with the gravitational fine structure constant $\alpha_\mathrm{G}=Gm_\mathrm{p}^2/\hbar c=m_\mathrm{p}^2/m_\mathrm{Pl}^2\cong 5\cdot 10^{-39}$. 

Inserting the latest CODATA values\cite{2018CODATA} of the proton mass and charge radius
leads to $n_\mathrm{c}=4.00087\cong 4$. Using Planck units with $h$ yields
$n_\mathrm{c}=0.636759\cong 2/\pi$.
This apparently geometric factor is assumed to be exactly $n_c\equiv 4$ allowing for a precise definition of the proton radius ($\hbar$ and $c$ are exact, and $m_\mathrm{p}$ is known with 11 digits),
\begin{equation}\label{RadiusProton}
r_\mathrm{p}
\equiv 4\cdot \frac{\hbar}{m_\mathrm{p}\cdot c}
=0.84123564119\cdot 10^{-15}~\mathrm{m}
\end{equation}
Comparing this value with the CODATA 2018 value, $r_\mathrm{p}=0.8414\,\pm\,0.0019$ fm, shows a deviation of only $\sim$218 ppm consistent with the standard uncertainty interval. Before the significant 2018 revision of $r_\mathrm{p}$, many attempts to resolve the large number puzzle were misguided by the measurements.
In future CODATA revisions, a slightly smaller proton radius is expected.

The definition in equation (\ref{RadiusProton}) is identical to the result of Trinhammer and Bohr \cite{TrinhammerBohr2019} which has a different approach. 

The remarkably simple relationship between the proton radius and mass by a factor of 4, 
corresponding to $n_\mathrm{c} = 4$ in equation (\ref{RadiusMass1019}), was associated with the trace anomaly in the energy momentum tensor in a heuristic QCD calculation\cite{XJi-1995}, where the author explained, ``3/4 of the nucleon mass comes from the traceless part of the energy-momentum tensor and 1/4 from the trace part. The magic number 4 is just the spacetime dimension.'' In a lecture note about the mass of baryonic matter\cite{XJi-Lecture}, the author elaborated with a calculation that minimized the total mass energy of a bag with three free quarks where the kinetic energy was $\propto 3/R$ and the energy of the free space was $\propto B(4\pi/3)\cdot R^3$. 
When the energy was minimized, the mass was exactly 4 times the kinetic energy of one quark: $M= 4/R$ (in suitable units).
A recent study builds on the result with lattice QCD calculations\cite{YB-Yang-2018}.

The micro cosmos model can be calibrated by requiring $m_\mathrm{mc}\equiv m_\mathrm{p}$ (using $m_\mathrm{p}$ as the input). Combining equations (\ref{MicroMassTotalInner}), (\ref{MicroMassTotalOuter}) and (\ref{RadiusProton}) gives:   
\begin{equation}\label{EqualityThreeExpressions}
\lambda_\mathrm{dB,p}
=\frac{h}{m_\mathrm{p}\cdot c}
=\frac{\pi}{2}\cdot r_\mathrm{p}
=2\pi\cdot\frac{l_\mathrm{Pl}\cdot m_\mathrm{Pl}}{m_\mathrm{p}}
=4\cdot a_\mathrm{max}
\end{equation}
or $r_\mathrm{p}=(8/\pi)\cdot a_\mathrm{max}$. Whenever the length ratio $a_\mathrm{max}/a_\mathrm{0}$ appears in the model, we can change to the mass ratio $m_\mathrm{Pl}/{m_\mathrm{p}}$, or vice versa, through the equality:
\begin{equation}\label{EqualityLargeNumbers}
\frac{a_\mathrm{max}}{a_0}
\equiv
\frac{\pi}{2}\cdot\frac{m_\mathrm{Pl}}{m_\mathrm{p}}
\end{equation}
where $m_\mathrm{Pl}/m_\mathrm{p}=1.3012\cdot 10^{19}$ and $a_\mathrm{max}=3.3035\cdot 10^{-16}$ m.
Equation (\ref{EqualityLargeNumbers}) can be used to quantify the predictions from the model to observable quantities.

The time scale $t_\mathrm{crunch}$ can be approximated by the photon crossing time
$t_\mathrm{crunch}\approx 2\cdot a_\mathrm{max}/c=2.2\cdot 10^{-24}$ s in agreement with the timescale of the strong interaction. The $\rho$ mesons have the shortest known lifetime, $\tau=4.5\cdot 10^{-24}$ s.

\section{Average nucleon density and time varying density}\label{SectionNucleonDensity}
An \emph{average nucleon density} can be defined as (inserting experimental values):
\begin{equation}\label{DensityNucleon}
\rho_\mathrm{nuc}
= \frac{m_\mathrm{p}}{\frac{4\pi}{3}\cdot r_\mathrm{p}^3}
=6.703\cdot 10^{17}~\mathrm{kg/m}^3
\end{equation}
Using the expression $r_\mathrm{p}=4\cdot l_\mathrm{Pl}\cdot m_\mathrm{Pl}/m_\mathrm{p}$ from equation (\ref{EqualityThreeExpressions}) and equations (\ref{RadiusMass1019})--(\ref{RadiusProton}), this is rewritten in close agreement with equation (\ref{DensityNucleon}):
\begin{eqnarray}\label{NuclearDensityM}
\rho_\mathrm{nuc}
&=&\frac{m_\mathrm{p}}
{\frac{4\pi}{3}\cdot \left(4\cdot\frac{l_\mathrm{Pl}\cdot m_\mathrm{Pl}}{m_\mathrm{p}}\right)^3}
=\frac{3}{4^4\cdot\pi}\cdot\left(\frac{m_\mathrm{p}}{m_\mathrm{Pl}}\right)^4\cdot\rho_\mathrm{Pl}
\end{eqnarray}
The relations $\rho=m/V\Leftrightarrow m=\rho\cdot V$ should only be expected to be valid when the quantities are integrated over one cycle.
We now define a time-varying density.
We cannot assume that the relations $\rho(t)=m(t)/V(t)\Leftrightarrow m(t)=\rho(t)\cdot V(t)$ are also valid for time-varying versions because, in general, the \emph{indefinite} integral of a fraction of two functions cannot be split into the fraction of the integrals of the two functions.
To counter this problem, we split the \emph{definite} integral while simultaneously introducing a calibration constant $k$ that depends on the fixed integration interval ($x$ is a time parameter running from 0 to 1):
\begin{equation}\label{Notes-IntegralSplit}
\int_0^1\rho(x)\,dx
=\int_0^1\frac{m(x)}{V(x)}\,dx
= k\cdot
\frac{\int_0^1 m(x)\,dx}
{\int_0^1 V(x)\,dx}
\end{equation}
It is now possible to define a \emph{time-varying density} expressed through $\rho_\mathrm{nuc}$:
\begin{equation}\label{NuclearDensityTimevarying}
\rho_\mathrm{nuc}(t)
=k\cdot\frac{m_\mathrm{mc}(t)}
{\frac{4\pi}{3}\cdot a(t)^3}
=k_\mathrm{c}\cdot\frac{a_\mathrm{max}^4}{a(t)^4}\cdot\rho_\mathrm{nuc}
\end{equation}
where $k_\mathrm{c}$ absorbed the constant $4\pi/3$ and it can be found by ensuring that:
\begin{eqnarray}\label{NuclearDensityIntegralCalibration}
\frac{1}{t_\mathrm{crunch}}\cdot\int^{t_\mathrm{crunch}}_0 \hspace{-3mm}\rho_\mathrm{nuc}(t)\,dt
=\rho_\mathrm{nuc}
\end{eqnarray}
Next we want to transform the time-varying density into a radial density.

\section{Radial density of the micro cosmos}\label{SectionRadialDensity}
\begin{figure}[t!]
\captionsetup{width=14cm}
\centerline{\includegraphics[width=14.0cm]{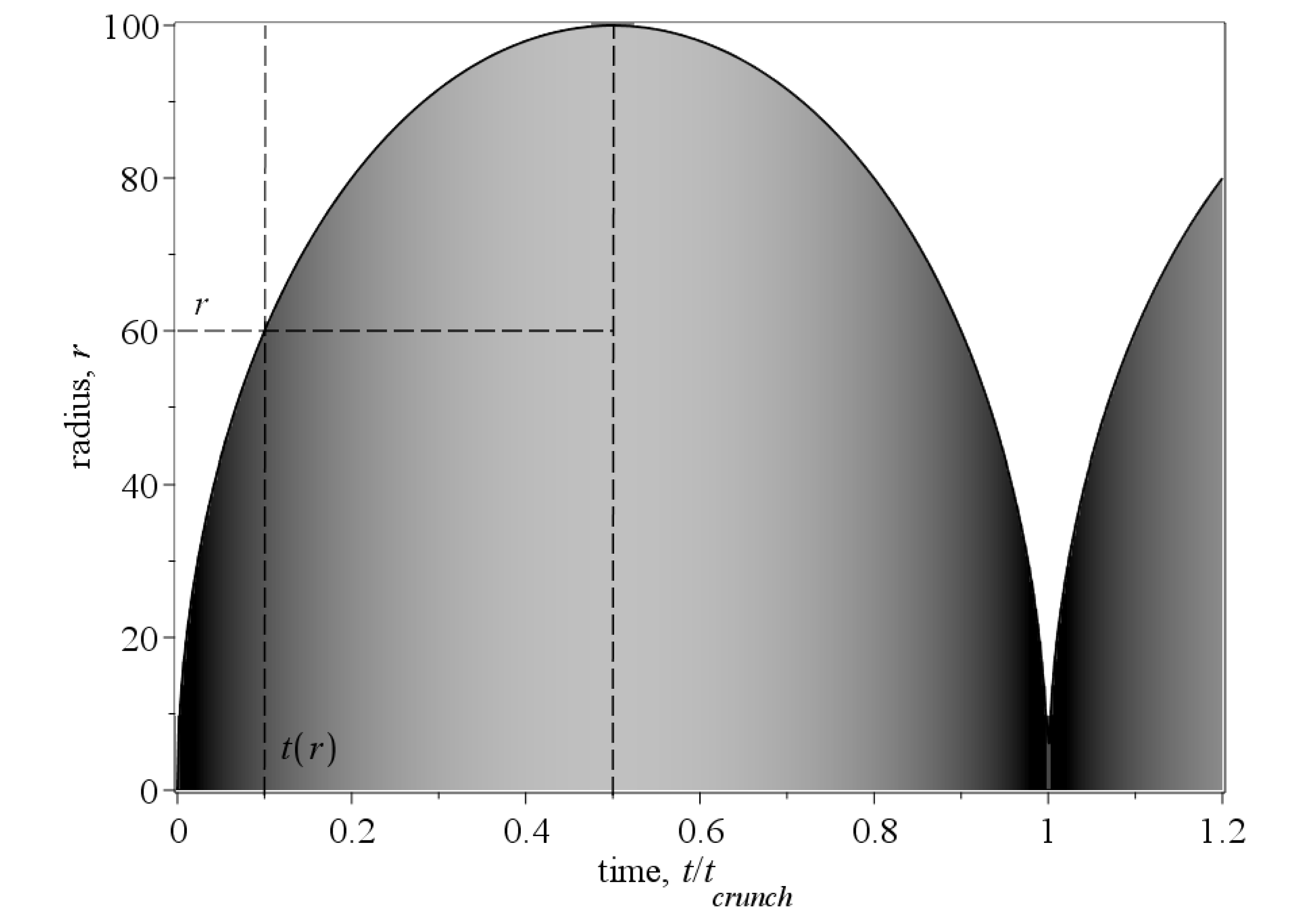}}
\vspace*{8pt}
\caption{The density is shown with grey scale shading where black is highest and light grey is lowest. The radial density $\rho(r)$ is found by integrating the horizontal dashed line from $t(r)$ to 1/2. Time is in units of $t_\mathrm{crunch}$. The values $a_\mathrm{max}=100$ and $a_\mathrm{0}=1$ are only chosen for illustration.}
\label{FigureRadialDensity}
\end{figure}

We seek a radial density profile $\rho(r)$ of the particle, which is calculated by integrating the density $\rho(t)$ over the relevant time interval, where the micro cosmos contributes to the density, and normalizing with this interval (see Fig.\ \ref{FigureRadialDensity}).

With equation (\ref{Time-scalefactor}),
$t(r)$\,=\,$t_\mathrm{crunch}\cdot\left(\frac{1}{2}-\frac{1}{2}\cdot\sqrt{(a_\mathrm{max}^2-r^2)/(a_\mathrm{max}^2-a_0^2)}\right)$,
where $a(t)$ is substituted with $r$, whereby the time variable is changed to a radial variable.
The integration interval is halved owing to the symmetry around $t=\frac{1}{2}t_\mathrm{crunch}$ so $t_1=t(r)$ and $t_2=\frac{1}{2}t_\mathrm{crunch}$.
The density in equation (\ref{NuclearDensityTimevarying}) is integrated and normalized to form the radial density as follows:
\begin{equation}\label{DefinitionRadialDensity}
\rho(r)
= k_\mathrm{c}\cdot
\frac{a_\mathrm{max}^4\cdot\rho_\mathrm{nuc}}{\frac{1}{2}\,t_\mathrm{crunch}-t(r)}\cdot
\int^{\frac{1}{2}\,t_\mathrm{crunch}}_{t(r)}
\frac{1}{a(t)^4}\,dt
\end{equation}
where the integral is written in appendix \ref{AppendixSectionRadialDensity}.
The calibration condition in equation (\ref{NuclearDensityIntegralCalibration}) is transformed to a condition for the total mass,
\begin{equation}
m_\mathrm{total}
=4\pi\cdot
\int_{a_0}^{a_\mathrm{max}} r^2\cdot\rho(r)\, dr
\equiv m_\mathrm{p}
\end{equation}
and the definition of $\rho_\mathrm{nuc}$ in equation (\ref{NuclearDensityM}) is used
leading to $k_\mathrm{c}=6\pi$ (while approximating the integral of the ln-term with a series expansion). The volume is rewritten, $a_\mathrm{max}^3=(\pi/8)^3\cdot r_p^3$, using equation (\ref{EqualityThreeExpressions}).
Inserting everything into equation (\ref{DefinitionRadialDensity}) gives the radial density profile averaged over one cycle time:
\begin{equation}
\rho(r)=3\pi
\cdot\left(\frac{a_\mathrm{max}^2}{r^2}
-\frac{1}{2}\cdot\sqrt{\frac{a_\mathrm{max}^2-a_0^2}{a_\mathrm{max}^2-r^2}}\cdot
\ln\left(\frac{a_\mathrm{max}+\sqrt{a_\mathrm{max}^2-r^2}}
{a_\mathrm{max}-\sqrt{a_\mathrm{max}^2-r^2}}\right)
\right)\cdot\rho_\mathrm{nuc}
\label{Radial-density-Master}
\end{equation}

\begin{figure}[t!]
\captionsetup{width=14cm}
\centerline{\includegraphics[width=14.0cm]{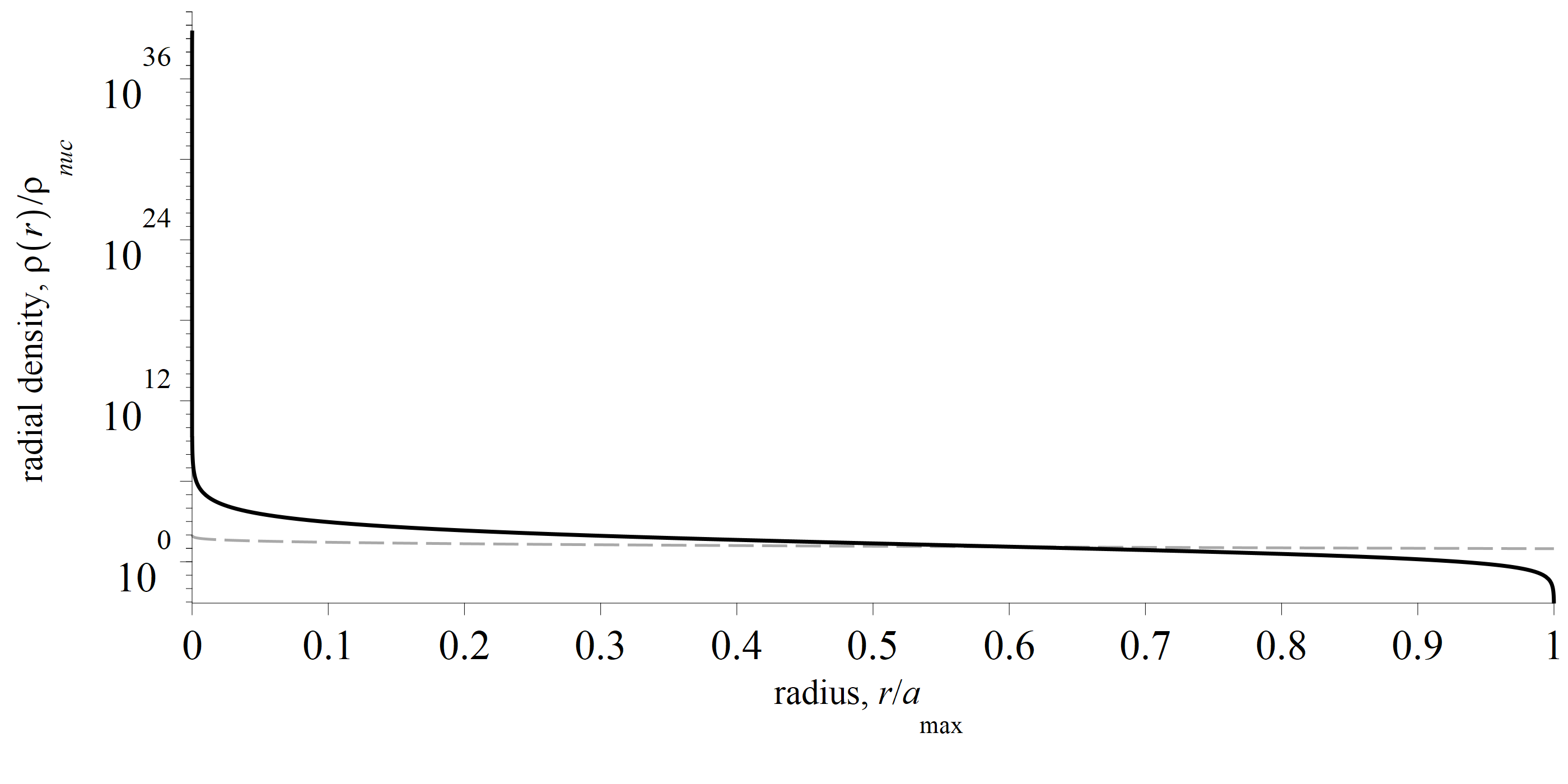}}
\vspace*{-4mm}
\caption{The radial density profile (solid black) consists of two terms. The $1/r^2$-term dominates in the center
approaching $4\cdot 10^{39}\cdot\rho_\mathrm{nuc}$ for $r\to a_0$. The ln-term shown in negative (dashed grey) is responsible for the cut-off near $r=a_\mathrm{max}$. The average nuclear density $\rho_\mathrm{nuc}$ corresponds to 1.}
\label{FigureRadialDensityProfile}
\end{figure}

The radial density is plotted in Fig.\ \ref{FigureRadialDensityProfile} showing a ``shoulder'' close to $r=a_\mathrm{max}$. Moving towards the center the dominating $1/r^2$-term approaches $4\cdot 10^{39}$ times the average nuclear density. This is surprisingly high but still much smaller than the Planck density.
It is expected that the central density is softened and broadened out in a generalization to an anisotropic model where the single point-like phase is split into three points of concentration (quarks), one for each dimension. 
In recent measurements\cite{Duran2023} the proton's ``mass radius'', which is dominated by the energy carried by gluons, was found to be smaller than the electric charge radius. This supports the existence of a dense core in the proton.

\section{Implications for nuclear physics}\label{SectionNuclearPhysics}
It is desirable to find a relationship between the radial density and the nucleon \emph{structure function}, which is used to describe deep inelastic scattering experiments.
A better understanding of the inner structure of a nucleon is important for nuclear theory.
The structure function is difficult to calculate exactly. It is typically assumed that the energy and charge densities follow each other. 
The momentum transfer is proportional to the radial velocity which in our model is equal to the time derivative of the scale factor: $v_r=da(t)/dt$.
The ``shoulder'' at large radius, due to the ln-term of equation (\ref{Radial-density-Master}), survives in the structure function as seen in experiments.

An interesting application would be to define a \emph{nucleon potential} (from $g_{00}$) and examine whether two micro cosmoses would bind and form stable systems (nuclei) or whether they would merge together or fly apart.
If each micro cosmos is forced to increase in size and cycle time when they are bound, the mass decreases, which is consistent with the \emph{mass defect} observed in bound nucleons.
The model has the properties required to describe the mass defect, and it might provide an alternative to models of nuclei that combine the methods in nuclear theory and QCD\cite{Miller2019} where it was found that a bound proton is larger than a free proton.

A radiation-dominated micro cosmos is fundamentally described as a ``perfect fluid'', so a liquid drop model seems appropriate phenomenologically. It is expected that the main terms of the binding energy, representing the volume and the surface of the liquid drop, can be described by the model.

The theory of graviweak unification mentioned in the introduction has been applied to calculations of the mass of atomic nuclei\cite{Kelabi-Elhmassi2023}, by adding a new term for the gravitational binding energy: 
$E_\mathrm{g}=-a_\mathrm{g}\cdot A(A-1)/A^{1/3}$ using the renormalized $G_F$. However we try to avoid the renormalization of coupling constants and instead connect the nucleon scale with the Planck scale via a micro cosmos.
A general relativistic treatment of quantum particles, such as that in our model, is preferable. Another example is light-front holographic QCD, which has been applied to the internal structure of hadrons such as pions and protons.\cite{TeramondEtal2021}

\section{A model of the electron}
\label{SectionElectron}
If we set the mass of the micro cosmos equal to the mass of an electron we  find (following Section \ref{SectionMassMicroCosmos}), the Compton wavelength $\lambda_\mathrm{C,e}=\hbar/(m_\mathrm{e}\cdot c)= 3.8616\cdot 10^{-13}$ m and the de Broglie wavelength $\lambda_\mathrm{dB,e}
=h/(m_\mathrm{e}\cdot c)=2.4263\cdot 10^{-12}$ m.
Both of these lengths are much larger than the generally accepted limit of $r_e\lesssim 10^{-18}~\mathrm{m}$
(experiments with a single electron in a Penning trap set an even smaller upper limit of the radius to $10^{-22}$ m).
The classical electron radius which links the electrostatic self-interaction energy to the relativistic mass-energy (assuming a constant charge density) is closer to the experimental limit:
\begin{equation}\label{ClassicalElectron}
r_e
=\frac{1}{4\pi\varepsilon_0}\cdot \frac{e^2}{m_\mathrm{e}\cdot c^2}
=\alpha\cdot \lambda_\mathrm{C,e}
=2.8179\cdot 10^{-15}~\mathrm{m}
\end{equation}
According to our model, the micro cosmos oscillates from the Planck length to the Compton length, and it is thus embracing both the classical radius and the experimental limits. The prediction that the density changes with time demands a calculation of a revised classical electron radius taking this into account. 

In the Kerr-Newman-Dirac model, the electron charge is assumed to be distributed on the surface of a disk-like oblate ellipsoid with thickness $r_e$ and radius $\lambda_{C,e}$, which depend also on the rotation parameter\cite{Lopez1984,Burinskii2015}. The two electron radii are thus connected geometrically via $\alpha$. On the border of the ellipsoid, a string-like structure (a naked ring singularity) is formed. From this ring, a cloud of virtual photons emanates. On the ring, one can associate ``traveling-wave excitations'' with quarks\cite{Burinskii2015}.
If a Dirac-Kerr-Newman black hole is embedded in a radiation-dominated micro cosmos, the radiation could (in the transition between two cycles) \emph{provide the source}, which has been a big mystery\cite{Burinskii2015} in the two-sheet version of the Kerr metric (representing a wormhole). The source has been suggested to be disk-like, shell-like, bag-like, and string-like\cite{Dymnikova2015}. A combination of Kerr-Newman-Dirac models could incorporate both the quarks and the gluon field of a proton.

\section{A micro cosmos with a point-like phase during the bounces}\label{SectionSingularity}
\begin{figure}[h!]
\captionsetup{width=14cm}
\hspace{2mm}
\centerline{\includegraphics[width=14cm]{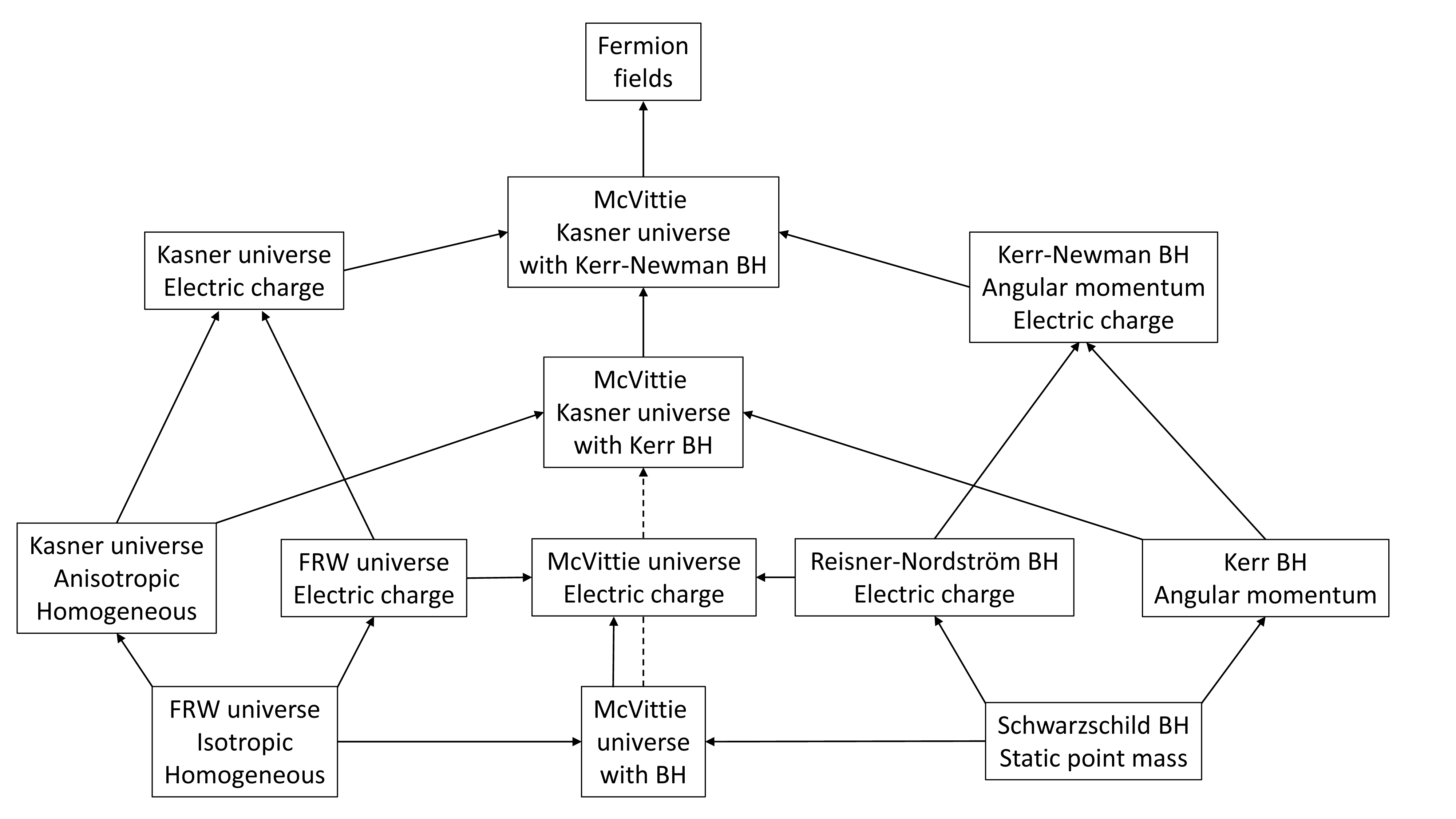}}
\caption{Diagram of metrics for model universes with or without isotropy merged with a black hole without or with electrical charge and angular momentum. The Dirac-Kerr-Newman electron corresponds to the box in the upper right corner. The complexity increases upwards and the built-in symmetries approach those describing fermion fields.}
\label{Figure-Diagram-Metrics}
\end{figure}

An external observer may describe the momentary point-like phase during the bounces as a micro-black hole.
The Schwarzschild metric describes the curved spacetime around a static point-like mass with mass $M$, parameterized by the Schwarzschild radius $r_s=2GM/c^2=2(M/m_\mathrm{Pl})\cdot l_\mathrm{Pl}$.
This model can be used as an approximation in small time intervals close to the bounces.
Inserting $M=m_\mathrm{mc}(t)=l_\mathrm{Pl}\cdot m_\mathrm{Pl}/a(t)$, from equation (\ref{MicroMassTimevarying}), in $r_s$, the mass is geometrized and depends only on $a(t)$ and the Planck length:
\begin{equation}\label{BH-mass}
r_s(t)=\frac{2\cdot M}{m_\mathrm{Pl}}\cdot l_\mathrm{Pl}
=\frac{2\cdot l_\mathrm{Pl}^2}{a(t)}
\end{equation}
Astrophysical black holes have $M\gg m_\mathrm{Pl}$ with a horizon $r_s\gg l_\mathrm{Pl}$. In our micro cosmos model $m_p\ll m_\mathrm{Pl}$ and we still have a sub-Planck scale $r_s$. However, the size of the particle is determined by \emph{the largest horizon}.
The horizon of the collapsing micro cosmos transforms into an expanding horizon of the black hole (they are inversely proportional). When the two horizons are equal, they interchange, and the micro cosmos turns inside-out. Assuming that there is no real singularity during the bounces\cite{WilsonEwing2018}, a singularity is avoided, and the next cycle begins.
An external observer observes that $a(t)\geq l_\mathrm{Pl}$ during a bounce. This dissolves the problem with the black hole horizon becoming smaller than the particle, and even smaller than the Planck length. A logical development of the model is to formally incorporate the momentary point-like phase during the bounces in a McVittie-like metric, as shown in the lower row of Fig.\ \ref{Figure-Diagram-Metrics}. 

The McVittie metric was applied by Faraoni and Jacques\cite{FaraoniJacques2007} to a larger system where the combined 'mass and scale factor' metric parameter, $m(t)=m_\mathrm{H}/a(t)$, is time-varying and parameterized through a quasilocal mass, $m_\mathrm{H}$, and the scale factor.
Our model is in accordance with this if $m_\mathrm{mc}(t)$ is identified with the $m$-parameter. During the phase with a nearly constant large $a\approx a_\mathrm{max}$ we have $m\approx m_\mathrm{particle}$. When $a(t)$ is small, the parameter $m$ increases accordingly.

A special case of the (closed) McVittie metric is the Kottler or Schwarzschild-de Sitter metric\cite{Kottler1918,Gaztanaka-MNRAS-2021,Gaztanaka-Symmetry-2022} which is characterized by two horizons $r_s$ and $r_H=c/H_\Lambda$:
\begin{equation}\label{SW-dS-metric-H}
ds^2 = -\left(1-\frac{r_s}{r}-(H_\Lambda/c)^2 r^2\right)\cdot c^2 dt^2
+\frac{1}{1-\frac{r_s}{r}-(H_\Lambda/c)^2 r^2}\,dr^2
+r^2d\Omega^2
\end{equation}
The metric reduces to the Schwarzschild metric when $H_\Lambda=0$. When $M=0\Rightarrow r_s=0$ it reduces to the de Sitter metric with static $H$ (omitting the index $\Lambda$ on $H$):   
\begin{equation}\label{Static-dS-metric}
ds^2 = -(1-(H/c)^2r^2)\cdot c^2 dt^2 
+\frac{1}{1-(H/c)^2r^2}dr^2 +r^2d\Omega^2
\end{equation}
The FLRW metric can be transformed to this by a suitable time variable (ensuring that the time and space derivatives of $T_{\mu\nu}$ are related by the factor $a\cdot H\cdot r$)\cite{Gaztanaka-MNRAS-2021,Gaztanaka-Symmetry-2022}.

\begin{figure}[h!]
\captionsetup{width=14cm}
\begin{center}
\mbox{\includegraphics[width=14cm]{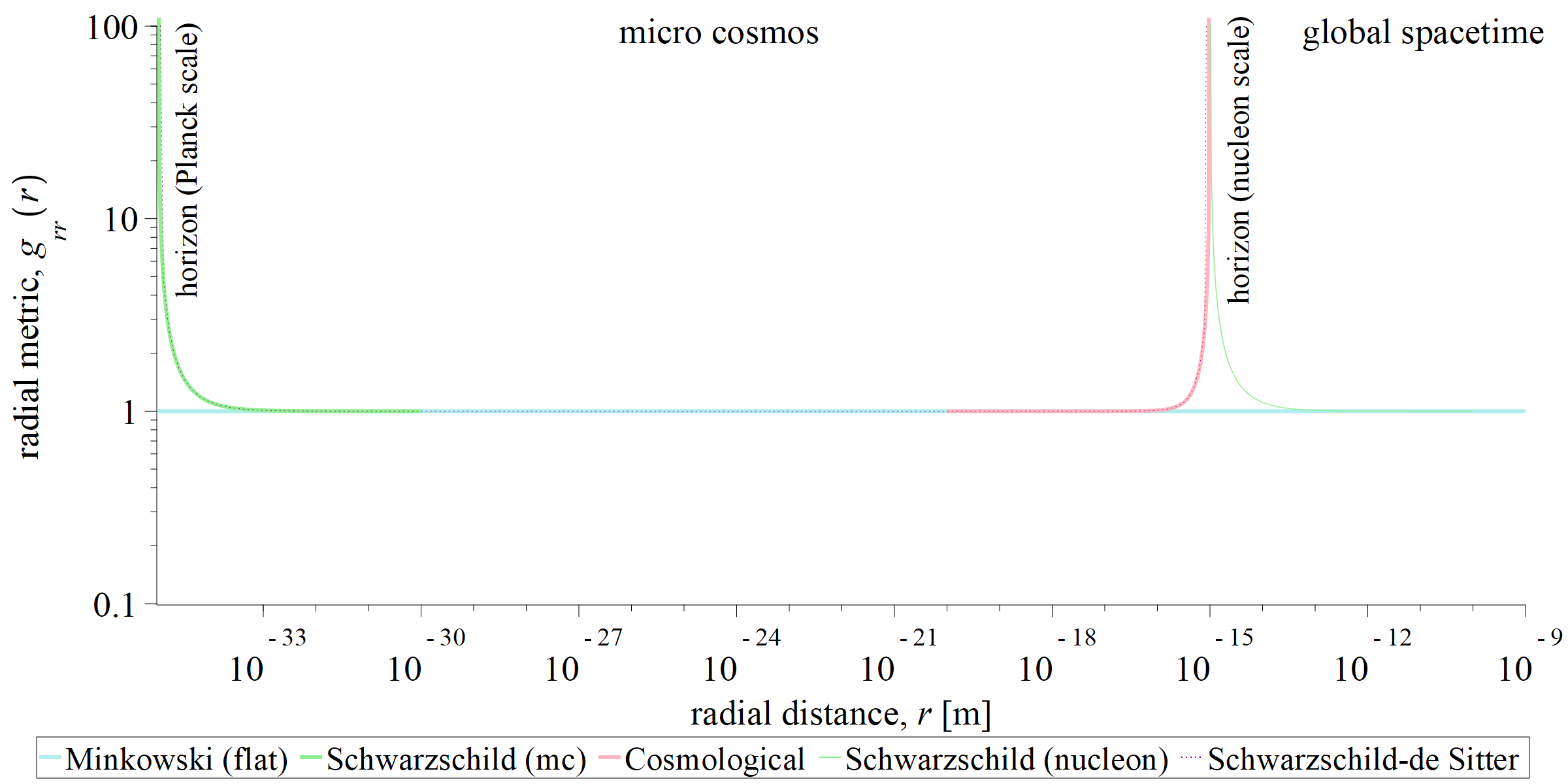}}
\end{center}
\vspace{-2mm}
\caption{Two Schwarzschild solutions (green) for $r_s=10^{-35}$ m (Planck length) and $r_s=10^{-15}$ m (nucleon); a cosmological solution (red) with $r_H=10^{-15}$ m together with the Schwarzschild-de Sitter solution (purple dotted curve) with $r_s=10^{-35}$ m and $r_H=10^{-15}$ m. All curves have the Minkowski background (light blue) as limits far from the horizons which are marked by text.}
\label{Figure-Radial-metric-Cosmo-Schwarz}
\end{figure}

The interior of a micro cosmos can be described by the Schwarzschild-de Sitter solution from $l_\mathrm{Pl}$ to $r_\mathrm{nuc}$. Figure \ref{Figure-Radial-metric-Cosmo-Schwarz} shows the radial component, $g_{rr}$, of the metric and $g_{tt}=-c^2\cdot g_{rr}^{-1}$. The nucleon horizon at $r=10^{-15}$ m is observed as a cosmological horizon from the inside and an event horizon from the outside. 
The next question is how the solution is continued through the horizon at $r=r_\mathrm{nuc}=10^{-15}$ m.

\section{The problem of embedding}
\label{SectionEmbedding}
It is important to consider how a particle described by a micro cosmos can fit into a background metric and move through space while still satisfying Einstein's field equations. A procedure to handle singularities by embedding was described in 1938 by Einstein, Infeld, and Hoffmann\cite{Einstein-Infeld-Hoffmann1938}. The embedding procedure is:
\vspace{-1.5mm}
\begin{itemize}
\item Linearization and superposition
\item Embedding and analytical continuation
\item Vanishing of the surface integral conditions
\end{itemize}
\vspace{-1.5mm}
The idea that the same horizon can be observed as a cosmological horizon from inside and an event horizon from outside was described earlier\cite{Frolov1990,Dymnikova2001} and it may allow a rich structure of baby universes embedded inside each other. It was also discussed recently\cite{Gaztanaka-MNRAS-2021} in an attempt to explain large-scale anomalies in the cosmic microwave background radiation. Here this matching is used to embed a micro cosmos in a black hole focusing on the radial component of the metric, $g_{rr}$, while $g_{tt}=-c^2\cdot g_{rr}^{-1}$.

\subsection*{\emph{Linearization and superposition}} 
The metric is split, $g_{\mu\nu}=\eta_{\mu\nu}+h_{\mu\nu}$, in a flat background, $\eta_{\mu\nu}$, and the deviation from the flat background is expanded by a power series: $h_{\mu\nu}=\sum_{l=1}^\infty \lambda^l(h_{\mu\nu})_l$, assuming that $h_{\mu\nu}$ depends continuously on a parameter $\lambda>0$ where the value $\lambda=0$ makes $h_{\mu\nu}$ vanish. The index $l$ on $h_{\mu\nu}$ is used for bookkeeping of the different nonlinear terms. The curvature tensor can also be expanded, $R_{\mu\nu}=\sum_{l=1}^\infty \lambda^l(R_{\mu\nu})_l=0$, and ordered in terms that are linear in $h_{\mu\nu}$, quadratic, and of higher order enabling linear superposition while satisfying the field equations\cite{Einstein-Infeld-Hoffmann1938}.

The two radial metric functions to be matched are the cosmological (cos) and the Schwarzschild (schw) $g_{rr}$, which are the limits of the Schwarzschild-de Sitter metric. They are written as series expansions at $r=0$ (cos), or $r=-\infty$, and $r=+\infty$ (schw):
\begin{eqnarray}\label{FLRW-SW-Grr-SeriesExpansions}
g_{rr}^\mathrm{cos}&=&
\frac{1}{1-r^2/r_H^2}
=\sum_{i=0}^{\infty}\left(\frac{r}{r_H}\right)^{2i}
= 1+\frac{r^2}{r_H^2}+\frac{r^4}{r_H^4}+\frac{r^6}{r_H^6}+\ldots\\
g_{rr}^\mathrm{schw}&=&
\frac{1}{1-r_s/r}
=\sum_{i=0}^{\infty}\left(\frac{r_s}{r}\right)^{i}
= 1+\frac{r_s}{r}+\frac{r_s^2}{r^2}+\frac{r_s^3}{r^3}+\ldots
\end{eqnarray}
The coinciding horizon, $r_0\equiv r_s=r_H$, and the radial coordinate functions are defined, $\lambda_H=r/r_H=r/r_0$ and $\lambda_s=r_s/r=r_0/r$, so $\lambda_H\cdot\lambda_s=1$.
If we define $\lambda\equiv\lambda_H=1/\lambda_s$ we ensure that the radial coordinate runs in the same direction and $\lambda=0$ makes $h_{\mu\nu}$ vanish.
A superposition of the two metrics (only relevant in $r=r_0$) can be constructed as a sum of powers of $\lambda(r)$ times $\eta_{rr}=1$, where the merged metric approaches the flat background at infinity:
$g_{\mu\nu}^\mathrm{merged}\to \eta_{\mu\nu}~\mathrm{for}~|r|\to\infty$. The metric is written: $g_{rr}=\eta_{rr}+h_{rr}^\mathrm{cos}+h_{rr}^\mathrm{schw}$ with $h_{rr}^\mathrm{cos}=
\sum_{i=0}^{N}\lambda^{2i}$ and $h_{rr}^\mathrm{schw}=
\sum_{i=0}^{N}\lambda^{i}$, where both sums are truncated to $N$ terms and the first term ($i=0$) is omitted (avoiding adding 1 twice).

\subsection*{\emph{Embedding and analytical continuation}} 
The central region behind an event horizon is embedded by a volume with a shell with thickness $\varepsilon$ which approaches zero in the matching procedure. 
In our model, the interior of the micro cosmos has to be matched with the exterior spacetime through the horizon (a coordinate singularity) by analytical continuation.
The function to be constructed must change smoothly through the shell ensuring it is continuous, without singularities, and differentiable to second order from minus to plus infinity.

\begin{figure}[h!]
\captionsetup{width=14cm}
\begin{center}
\mbox{\includegraphics[width=14cm]{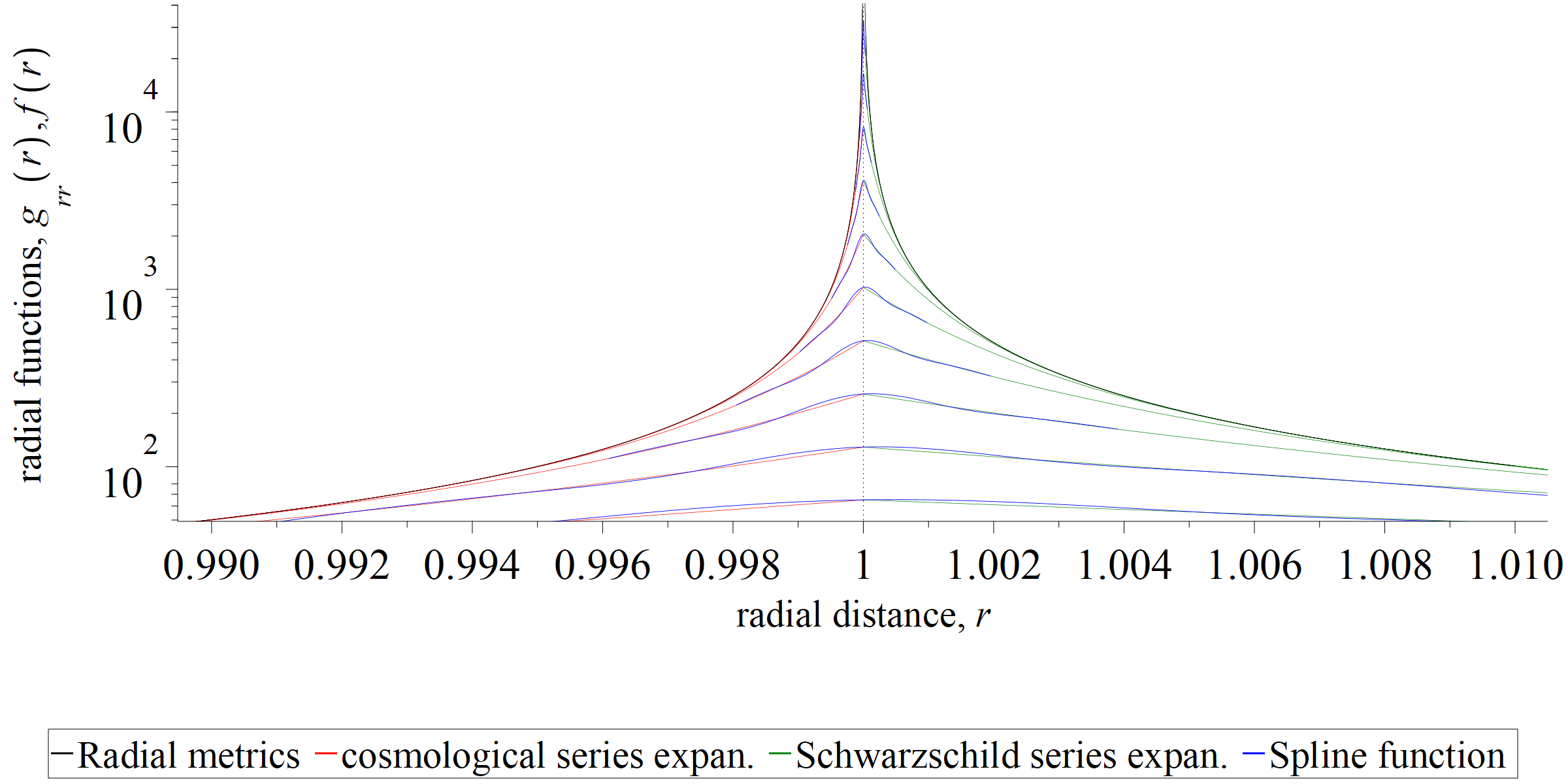}}
\end{center}
\vspace{-2mm}
\caption{Log-plot of series expansions to cosmological solutions (red) and Schwarzschild solutions (green) both functions with the horizon at $r_0=1$ (vertical black dotted line) and both approaching the original $g_{rr}$ functions (black curves). The cubic spline interpolation functions (blue) are continuing the two curves through the horizon. The curves have $N=2^m$ terms for $m=6..15$ (up to 32768 terms) and the width of the interval $r_0\pm\varepsilon$ is decreasing with increasing $N$.}
\label{Figure-Radial-Spline-Zoom}
\end{figure}

A very powerful method of constructing an analytical function $S(r)$ that interpolates a series of data points is the \emph{cubic spline interpolation method}. The setup of this scheme is described in \ref{AppendixSplineSetup} and the result is shown in Fig.\ \ref{Figure-Radial-Spline-Zoom}.

\subsection*{\emph{Vanishing of the surface integral conditions}} 
It is required by the field equations that any surface integral terms coming from the enclosing of a region should vanish\cite{Einstein-Infeld-Hoffmann1938}.

The two series coincide at the horizon $g_{rr}^\mathrm{cos}=g_{rr}^\mathrm{schw}=N+1$ for an arbitrary number of terms $N<\infty$.
We can choose that $\varepsilon$ decreases as $\varepsilon(N)=r_0/N$ (see appendix \ref{AppendixSplineSetup}). 
When $\varepsilon\to 0$ the endpoints (and all intermediate points) are drawn towards $r_0$ and the spline interpolation function approaches a Dirac delta function. This is realized by a Gauss function with finite maximum, $\delta(r_0)=N+1$, and standard deviation, $\sigma=\varepsilon/\sqrt{2\ln(2)}=1/(N\sqrt{2\ln(2)})$:
$\delta(r-r_0)=(N+1)\cdot\exp(-\ln(2)N^2(r-r_0)^2)$.

When the embedding surface shrinks $S(r_0)\to\delta(r_0)$
and any functions depending on the coordinate $r_0$ go over to a sum of Dirac delta functions, thus avoiding infinite values in the analytical continuation of the field.

\section{Perspective for future generalizations}\label{SectionGeneralization}
Future generalizations could include the following challenges:
1) The electrical charge may be included in the metric for a Reissner-Nordström black hole in a FLRW background\cite{McVittieCharge};
2) The point-like phase may also account for angular momentum (spin) with the Kerr metric, or both electrical charge and spin with the Kerr-Newman metric, enabling a description of the difference between matter and antimatter;
3) The extended phase of an anisotropic micro cosmos may be described by the Kasner metric, providing external/geometric degrees of freedom that can be associated with intrinsic quantum numbers such as spin and charges;
4) The Kasner and Kerr-Newman metrics may be combined into a general McVittie-like metric with sufficiently high complexity to describe the fermion fields, as shown in Fig.\ \ref{Figure-Diagram-Metrics};
5) The electromagnetic radiation assumed in the model may be substituted with a gluon equation of state (the photon has total 'white' color charge) and
6) Leptons and quarks may be modeled with another (sub 3d) geometry.

An additional complication to the inclusion of quantum properties (in the point-like phase) is how the different geometries should be embedded within each other. 

\section{Summary of the results}
A quantum particle was modeled using a standard radiation-dominated homogeneous and isotropic FLRW model, where the usual parameters $\Omega_0$ and $H_0$ were substituted with the maximal size $a_\mathrm{max}$ and cycle time $t_\mathrm{crunch}$, emphasizing the geometric nature of the model. The Friedmann equation was solved for the scale factor. 

The model is rooted in the original field equations and the Planck scale with no change in the coupling constant (as in graviweak unification and strong gravity), making the model simple and robust.

The mass of the particle was calculated and found to vary with time.
The total mass integrated over one cycle was determined and identified with the particle mass measured in experiments.
An inverse proportionality between the size and mass of the particle was established to be identical to the Compton wavelength (for the time-varying mass) and the de Broglie wavelength (for the total mass).

The model was applied to the proton, and a relationship between the proton radius and proton mass was determined in agreement with experiments.
The radial density profile of the nucleon was calculated, and found to exhibit a shoulder, which is a characteristic of the nucleon structure function and the density increases towards the center.
The model has the properties required to describe the binding between nucleons and the mass defect in atomic nuclei.

The micro cosmos momentarily acts as a black hole during the bounces. The largest horizon defines the size of the particle, thus avoiding sub-Planck distances. The Kottler-Schwarzschild-de Sitter metric describes both phases.

\appendix
\numberwithin{equation}{section}


\section{Integration of the mass}
\label{AppendixIntegrationMass}
The integral of the inverse scale factor is a standard integral in $x=t/t_\mathrm{crunch}$ with $A=a_0^2$, $B=4\cdot(a_\mathrm{max}^2-a_0^2)\cong 4\cdot a_\mathrm{max}^2\,\Rightarrow\,\sqrt{B}\cong 2\cdot a_\mathrm{max}$ and $C=-B$. The term $4A/B\cong a_0^2/a_\mathrm{max}^2\ll 1$ is omitted in the square root in the denominator of arcsin: 
\begin{eqnarray}\label{IntegralInverseScaleFactor1}
\int
\frac{1}{a(x)}\,dx
&=&\int\frac{1}{\sqrt{A+B\cdot x+C\cdot x^2}}\,dx\\
&=&\frac{-1}{\sqrt{-C}}\cdot
\arcsin\left(\frac{2\cdot C\cdot x+B}{\sqrt{B^2-4\cdot A\cdot C}}\right)\\
&=&\frac{-1}{\sqrt{B}}\cdot
\arcsin\left(\frac{1-2\cdot x}{\sqrt{1+4\cdot A/B}}\right)\\
&=& -\frac{1}{2\cdot a_\mathrm{max}}\cdot
\arcsin\left(1-2\cdot x\right)
\end{eqnarray}
Inserting the limits gives:
\begin{eqnarray}
\int^{1}_{0} \frac{1}{a(x)}\,dx
&=&
-\frac{1}{2\cdot a_\mathrm{max}}\cdot
[\arcsin\left(1-2\cdot x\right)]^{1}_{0}
= \frac{\pi}{2}\cdot\frac{1}{a_\mathrm{max}}
\end{eqnarray}

\section{Radial density calculation}
\label{AppendixSectionRadialDensity}
The scale factor squared was written in standard form with $x=t/t_\mathrm{crunch}$ and $K=4\cdot a_\mathrm{max}^2$,
\begin{eqnarray}\label{Scale-factor-squared-rewritten}
a(x)^2
&=& a_0^2 +4\cdot(a_\mathrm{max}^2-a_0^2)\cdot x
-4\cdot(a_\mathrm{max}^2-a_0^2)\cdot x^2\\
&=& \left(\frac{1}{4}\cdot\frac{a_0^2}{a_\mathrm{max}^2}
+\left(1-\frac{a_0^2}{a_\mathrm{max}^2}\right)\cdot x
-\left(1-\frac{a_0^2}{a_\mathrm{max}^2}\right)\cdot x^2\right)
\cdot 4\cdot a_\mathrm{max}^2\\
&=& \left(A +(1-4A)\cdot x -(1-4A)\cdot x^2\right)\cdot K\\
&=& \left(A +B\cdot x +C\cdot x^2\right)\cdot K
\end{eqnarray}
where $B=1-4A$, $C=-(1-4A)$ and $A$ is rewritten with equation (\ref{EqualityLargeNumbers}):
\begin{equation}\label{ConstantA}
A=\frac{1}{4}\cdot\left(\frac{a_0}{a_\mathrm{max}}\right)^2
=\frac{1}{\pi^2}\cdot\left(\frac{m_\mathrm{p}}{m_\mathrm{Pl}}\right)^2
=5.9842\cdot 10^{-40}  
\end{equation}
The radial density in equation (\ref{DefinitionRadialDensity}) is rewritten using $x=t/t_\mathrm{crunch}$:
\begin{eqnarray}
\rho(r)
&=& k_\mathrm{c}\cdot
\frac{a_\mathrm{max}^4\cdot\rho_\mathrm{nuc}}{\frac{1}{2}\,t_\mathrm{crunch}-t(r)}\cdot
\int^{\frac{1}{2}\,t_\mathrm{crunch}}_{t(r)}
\frac{1}{a(t)^4}\,dt\\
&=& k_\mathrm{c}\cdot
\frac{a_\mathrm{max}^4\cdot\rho_\mathrm{nuc}\cdot t_\mathrm{crunch}}{\frac{1}{2}\,t_\mathrm{crunch}-x(r)\cdot t_\mathrm{crunch}}
\cdot\int^{\frac{1}{2}}_{x(r)}
\frac{1}{a(x)^4}\,dx\\
&=& \frac{k_\mathrm{c}}{16}\cdot
\frac{\rho_\mathrm{nuc}}{\frac{1}{2}-x(r)}
\cdot\int^{\frac{1}{2}}_{x(r)}
\frac{1}{(A +B\cdot x +C\cdot x^2)^2}\,dx
\label{Radial-density}
\end{eqnarray}
The indefinite integral is a standard integral:
\begin{eqnarray}
\int \frac{1}{(A+B\cdot x-C\cdot x^2)^2}\,dx
&=&\frac{2\cdot x-1}{A+(1-4A)\cdot x-(1-4A)\cdot x^2}\nonumber\\
&&+2\cdot
\ln\left(\frac{A+(1-2A)\cdot x}{1-A-(1-2A)\cdot x}\right)
\label{Density-Integral-Indefinite}
\end{eqnarray}
The two terms of the integral are integrated separately.
\begin{eqnarray}
\int^{\frac{1}{2}\,t_\mathrm{crunch}}_{t(r)}
\frac{1}{a(t)^4}\,dt
&=&\frac{t_\mathrm{crunch}}{4\cdot a_\mathrm{max}^4}\cdot
\sqrt{\frac{a_\mathrm{max}^2-r^2}{a_\mathrm{max}^2-a_0^2}}
\cdot\frac{a_\mathrm{max}^2}{r^2}\nonumber\\
&&-\frac{t_\mathrm{crunch}}{8\cdot a_\mathrm{max}^4}\cdot
\ln\left(
\frac{a_\mathrm{max}+\sqrt{a_\mathrm{max}^2-r^2}}
{a_\mathrm{max}-\sqrt{a_\mathrm{max}^2-r^2}}
\right)
\label{Radial-Density-Result}
\end{eqnarray}

\vspace{-5mm}
\section{Analytical continuation spline function setup}
\label{AppendixSplineSetup}
We choose a data point interval with 7 points centered around $r_0$ at the horizon:
\begin{equation}
[r_0-\varepsilon, r_0-\tfrac{2}{3}\varepsilon, r_0-\tfrac{1}{3}\varepsilon, r_0, 
r_0+\tfrac{1}{3}\varepsilon, r_0+\tfrac{2}{3}\varepsilon, r_0+\varepsilon]
\end{equation}
We also define the functions: 
\begin{eqnarray}
&&f_a(r)\equiv 1+h_{rr}^\mathrm{cos},
~f_a'(r)\equiv\partial_r h_{rr}^\mathrm{cos},
~ f_a''(r)\equiv\partial_r^2 h_{rr}^\mathrm{cos},\\ 
&&f_b(r)\equiv 1+h_{rr}^\mathrm{schw},
~f_b'(r)\equiv\partial_r h_{rr}^\mathrm{schw},
~f_b''(r)\equiv\partial_r^2 h_{rr}^\mathrm{schw}
\end{eqnarray}
A piecewise polynomial spline function $S(r)$ is constructed that matches the two functions in the endpoints:
\begin{equation}\label{SplineFunction}
g_{rr}(r)=\left\{
\begin{array}{ll}
f_a(r) & \mathrm{for}~ r\leq r_0-\varepsilon\\
S(r)   & \mathrm{for}~ r_0-\varepsilon\leq r\leq r_0+\varepsilon\\
h_b(r) & \mathrm{for}~ r\geq r_0-\varepsilon\\
\end{array}
\qquad
\right.
\end{equation}
In addition to ensuring continuity, $S(r)$ must also ensure that the first and second derivatives in the endpoints are equal to the derivatives of the functions:
\begin{eqnarray}
\mathrm{Continuity:~} && S(r_0-\varepsilon) = f_a(r_0-\varepsilon), \quad
S(r_0+\varepsilon) = f_b(r_0+\varepsilon)\\ 
\mathrm{1.~derivaties:} && S'(r_0-\varepsilon) = f_a'(r_0-\varepsilon), \quad
S'(r_0+\varepsilon) = f_b'(r_0+\varepsilon) \\ 
\mathrm{2.~derivaties:} && S''(r_0-\varepsilon) = f_a''(r_0-\varepsilon), \quad
S''(r_0+\varepsilon) = f_b''(r_0+\varepsilon)
\end{eqnarray}
The slopes of the two exterior functions get steeper and steeper
and the intersection angle between the curves becomes more and more acute.
To ensure that an analytical continuation is possible through the horizon, 
$\varepsilon$ must be decreased systematically with increasing $N$.
Inserting $r=r_0-\varepsilon$ in $g^\mathrm{cos}_{rr}$ and $r=r_0+\varepsilon$ in $g^\mathrm{schw}_{rr}$ and solving for $\varepsilon$ leads to the relations $\varepsilon_a(N)=r_0/(2N+2)$ and $\varepsilon_b(N)=r_0/N$. We can choose $\varepsilon(N)=r_0/N$ for both sides.

The set of equations can be solved with linear algebra and is programmed in most CAS software. The result is shown in Fig.\ \ref{Figure-Radial-Spline-Zoom}.

\section*{Acknowledgments}
The author wishes to thank Steen H. Hansen, Troels Harmark, Helge S.\ Kragh and Jesper Petersen for comments on earlier versions of the manuscript and Hans O.U.\ Fynbo for stimulating discussions about nuclear structure. 
The author also thanks the anonymous referees for their helpful comments that improved the quality of the manuscript.
This work has not enjoyed support by any grants. Christianshavns Gymnasium is thanked for providing time for professional upskilling along teaching obligations.
The author declares that no competing interests exist.

This manuscript is found with comments on the website \href{http://micro.cozmo.dk}{www.micro.cozmo.dk}.


\end{document}